\documentclass[12pt]{article}

\newcommand{\manuscript}{On nanostructured molybdenum-copper composites produced by high pressure torsion}
\newcommand{\shortmanuscript}{HPT of Mo-Cu}
\newcommand{\shortauthors}{Rosalie, Guo, Pippan, Zhang}
\usepackage[numbers,sort&compress]{natbib}

\usepackage[left=2.5cm,right=2.5cm,top=2cm,bottom=2cm]{geometry}
\usepackage{fancyhdr}
	\fancypagestyle{plain}{ %
	\fancyhf{} 
	\lhead{\textit{\shortmanuscript}}
	\rhead{\shortauthors}
	\lfoot{}
	\rfoot{\thepage}
	}

\usepackage{amsbsy} 
\usepackage{graphicx}
	\graphicspath{{./}{./Figures/}} 
	\DeclareGraphicsExtensions{.pdf,.png,.jpg}

\usepackage{subfigure}
\usepackage[version=3]{mhchem} 
\usepackage{multirow} 
\usepackage{multicol}
\usepackage{booktabs}	
\usepackage{setspace}
\usepackage{xspace}
\usepackage{comment}
\usepackage{amssymb}

\usepackage{doi} 

\title{\manuscript}

\author{Julian M. Rosalie$^{1,*}$,
\and Jinming Guo$^{1}$,
\and Reinhard Pippan$^{1}$,
\and Zaoli Zhang$^{1}$
}	
\date{}
\vspace{6pt}

\begin{document}

\maketitle
$^{1}$Erich Schmid Institute, Austrian Academy of Sciences, Jahnstrasse 12, Leoben, 8700, Austria.\\
\begin{abstract} 
Nano-structured molybdenum-copper composites have been produced through severe plastic deformation of liquid-metal infiltrated Cu30Mo70 and Cu50Mo50 (wt.\%) starting materials. Processing was carried out using high pressure torsion at room temperature with no subsequent sintering treatment, producing a porosity-free, ultrafine grained composite. Extensive deformation of the Cu50Mo50 composite via two-step high-pressure torsion produced equiaxed nanoscale grains of Mo and Cu with a grain size of 10--15\,nm. Identical treatment of Cu30Mo70 produced a ultrafine, lamellar structure, comprised of Cu and Mo layers with thicknesses of $\sim$5\,nm and $\sim10-20$\,nm, respectively and an interlamellar spacing of 9\,nm.  This microstructure differs substantially from that of HPT-deformed Cu-Cr and Cu-W composites, in which the lamellar microstructure breaks down at high strains. The ultrafine-grained structure and absence of porosity resulted in composites with Vickers hardness values of 600 for Cu30Mo70 and 475 for Cu50Mo50. The ability to produce Cu30Mo70  nanocomposites with a combination of high strength, and a fine, oriented microstructure should be of interest for thermoelectric applications. 
\end{abstract}


\section{Introduction}

Molybdenum-copper composites are used in high power electrical applications such as thermoelectrics where a combination of dimensional stability, high mechanical strength and high thermal and electrical conductivity are required \cite{AguilarGuzman2011,GuoKang2012,KumarJayasankar2015}. 
Molybdenum has the high melting point (2610$^\circ$C\cite{Cockeram2002}) and low thermal-expansion coefficient \cite{FangZhou2014} characteristic of refractory metals such as Nb, Ta, W and Cr. While molybdenum has similar electrical and thermal conductivity to tungsten, the lower density of Mo makes it preferred for weight-critical applications \cite{FangZhou2014} and its lower ductile-to-brittle-transformation temperature (DBTT) of $\sim$373\,K \cite{ShieldsLipetzky2001} makes forming Mo-Cu composites easier than equivalent Cu-W composites. In addition to providing the composites with thermal and electrical conductivity, copper also increases the ductility of the composites  \cite{FangZhou2014,FangZhou2013} which is useful both during production and in service.

Mo and Cu have a positive enthalpy  of mixing of +18\,Kj/mol and have negligible solid solubility even at the melting temperature of copper \cite{SubramanianLaughlin1990}.  The lack of mutual solubility is advantageous for retaining high conductivity \cite{KumarJayasankar2015,Friedel1956}, however, it also results in a weak interface and hence poor mechanical bonding between the phases\cite{RodriguezGoodman1992}. 

Mo-Cu composites are conventionally produced via liquid metal infiltration \cite{FangZhou2014,JohnsonGerman2001}. This involves sintering Mo powders to make a porous pre-form which can then be infiltrated with molten Cu. This process is only viable in a limited composition range, and the rate of densification is low\cite{WangLiang2015}.

Efforts to accelerate densification have mainly involved using ball milling to produce nanoscale Mo and Cu powders with high defect concentrations. These can then be densified more rapidly  via powder compaction and sintering at  1000--1200\,$^\circ$C \cite{WangLiang2015,TikhiiKachalin2007,XiZuo2008}. Although the ball-milled powders have grain sizes in the range of 10-20\,nm, grain growth  during sintering is rapid and the densified compacts have grain diameters of several microns\cite{KumarJayasankar2015,JohnsonGerman2001,TikhiiKachalin2007}. Complete densification remains unachievable and intergrannular porosity is still observed after sintering \cite{KumarJayasankar2015,TikhiiKachalin2007}.

High pressure torsion (HPT) has been used for grain refinement and mechanically alloying of a wide range of Cu-based systems including Cu-Fe\cite{MojtahediGoodarzi2013,SoleimanianMojtahedi2014,BachmaierKerber2012}, Cu-Cr \cite{BachmaierRathmayr2014,Islamgaliev2014}, Cu-Nb\cite{EkizLach2014,AbadParker2015} and Cu-W\cite{Raharijaona2011,KraemerWurster2014}. 
Deformation is carried out under quasi-hydrostatic conditions, allowing brittle materials such as refractory metals to be extensively strained without the need for external heating. This, in turn, avoids the grain growth which accompanies high temperature sintering. Previous studies on pure Mo found more extensive grain refinement could be achieved with HPT (with an average grain size of 0.2\,$\mu$m) than with either equal channel angular extrusion (ECAE) or multi-stage forming (average grain sizes of 2.5\,$\mu$m and 1.8\,$\mu$m, respectively) due to the lower processing temperature\cite{Ivanov2008}. HPT imposes a strain that is proportional to the radial distance at a given point and thus a single sample can provide information covering a wide range of strain conditions. 

The objective of this investigation was to use HPT to produce bulk, full-dense, nano-grained Mo-Cu composites and to examine their microstructures at different applied strains. Although there are number of reports on the HPT-deformation of the related Cu-Cr (see for example,  \cite{BachmaierRathmayr2014,Islamgaliev2014,SauvageJessner2008}) and several on Cu-W\cite{Raharijaona2011,KraemerWurster2014} systems, to the best of the authors' knowledge this work provides the first examination of HPT-deformed Cu-Mo composites.

\section{Experimental details}

This investigation was carried out using two liquid-metal infiltrated Mo-Cu composites provided by Plansee SE, Austria. These composites had compositions of Mo-30\,wt\%Cu (40\,at.\%Cu) and Mo-50\,wt.\%Cu (60\,at.\%Cu). These are hereafter described as Cu30Mo70 and Cu50Mo50, respectively. 

Deformation was applied using multi-step high-pressure torsion (HPT) \cite{BachmaierKerber2012}.
Discs (height, $h\sim12$\,mm, diameter, $\phi=30$\,mm) were cut from the raw ingots by electrical discharge machining. 
These were deformed at room temperature in a 4000\,kN HPT apparatus, using tool steel anvils with an anvil gap of 3.5\,mm. A total of either 5 or 10 complete rotations (hereafter described as $N_1$=5 or 10, with the subscript indicating the first step of deformation) were performed at an angular velocity ($\omega$) of 0.0625 revolutions per minute (rpm) under an applied pressure of $\sim$4.5\,GPa. 

\begin{figure*}[hbtp]
	\begin{center}
		\includegraphics[width=12cm]{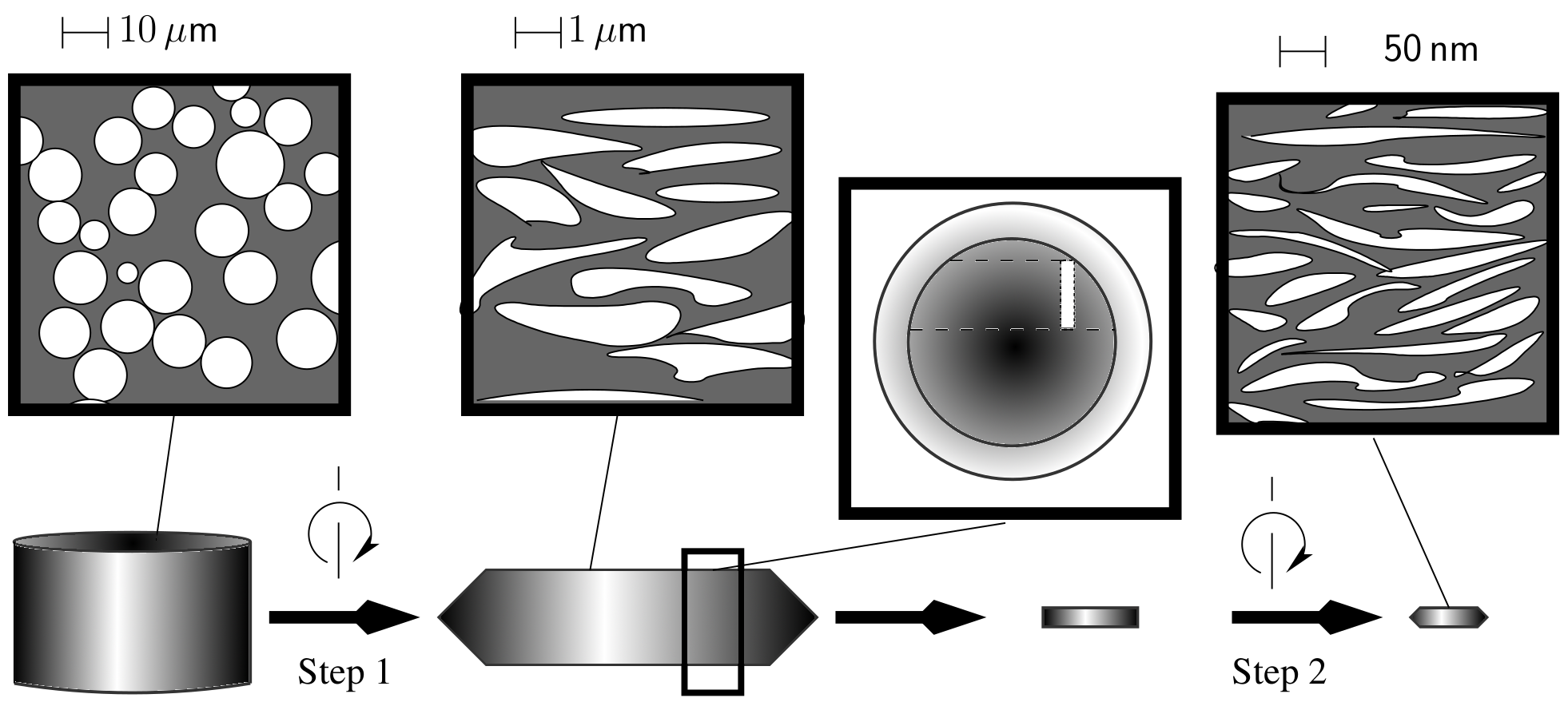}
	\caption{Schematic representation of the sample geometry and corresponding microstructure at each stage of processing. The microstructures produced by step one and step two HPT deformation are described in detail in the following section. Note that the shear plane during step 2 HPT is perpendicular to that of step one deformation. \label{fig-sample-geometry}}
	\end{center}
\end{figure*}

Cylindrical samples (h$\sim$20\,mm $\phi=7.5$\,mm) were cut from the discs, perpendicular to the shear plane of this disc (See Fig.~\ref{fig-sample-geometry}). These cylinders were sectioned into samples with height $\sim$1.1\,mm. The average strain in these discs was calculated as 25 for $N_1=5$ and 50 for $N_1=10$. Step two deformation was conducted using a 400\,kN HPT apparatus at an applied pressure of 7.4\,GPa and $\omega$=0.4\,rpm.  The second step of deformation consisted of $N_2=10-50$  rotations of deformation with a 0.25\,mm anvil gap and cavity diameter of 8\,mm.

Vickers micro-indentation hardness testing was conducted on as-received material and following each step of HPT deformation. 
Scanning electron microscopy was used to characterise the microstructure in the as-received condition and following HPT deformation. Samples for SEM examination were ground and polished using standard techniques, within final polishing taking place on a Buehler Vibromet polisher. SEM examinations were carried out using a Zeiss Leo 1525 microscope equipped with field emission gun and Gemini lens system. Micrographs were recorded using the backscattered electron detector (BSD) with source voltages of 15--20\,kV. Stereological measurements were made using ImageJ software (Version 1.50g). The Mo particle size in the as-received condition was measured by a processing consisting of (i) background subtraction (ii) thresholding (iii) separation of connected particles via the watershed algorithm and (iv) measurement of the projected particle area. Determining the lamellar spacing required similar preprocessing steps of background subtraction and thresholding. The spacing was then measured by counting the number of intercepts using a circular grid to give the true mean spacing\cite{VanderVoort1984}. This follows the established methodology developed for measuring lamellar spacings in pearlitic steels.

\begin{figure*}[hbtp]
	\begin{center}
		\includegraphics[width=12cm]{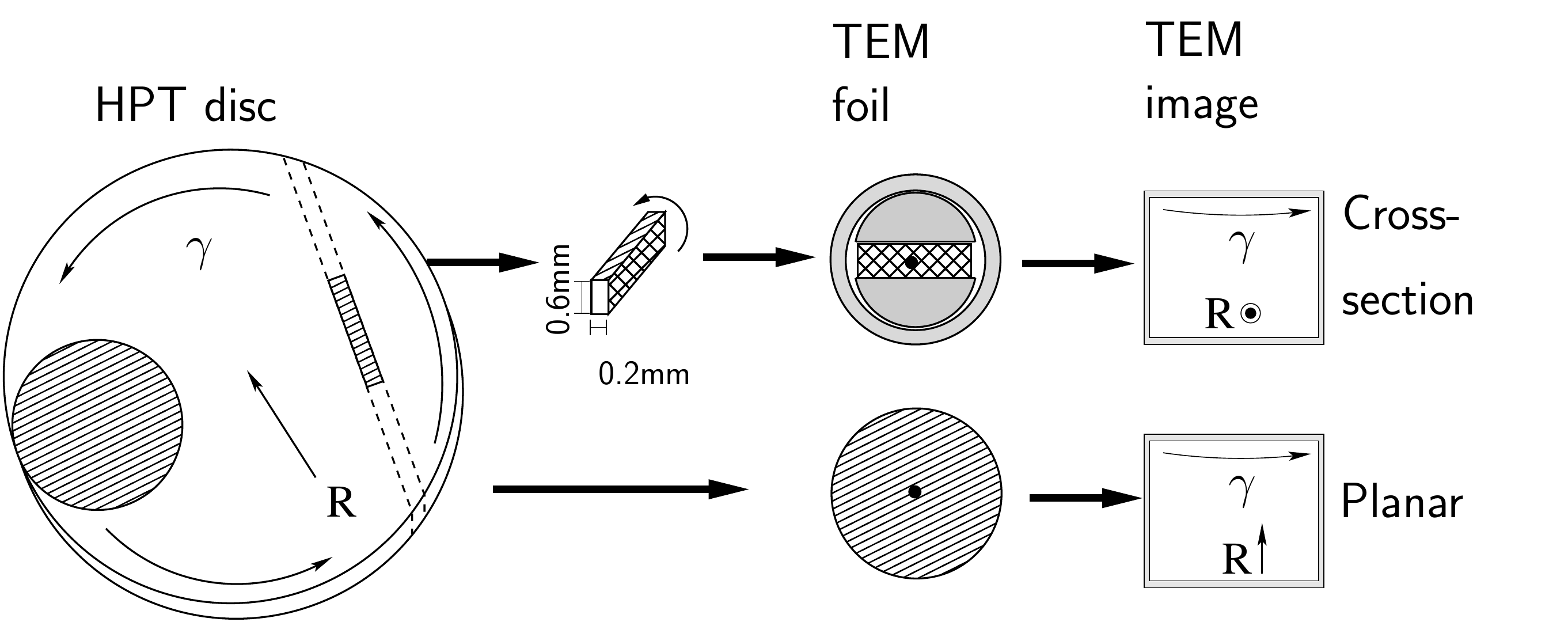}
	\caption{Schematic showing the relationship between the HPT disc and the orientation of the planar and cross-sectional TEM samples.  For both sample types the shear direction is normal to the electron beam.\label{fig-foil-geometry}}
	\end{center}
\end{figure*}

Cross-sectional and plan-section TEM foils prepared by dimple grinding to a thickness of 10-15\,$\mu$m, followed by precision ion polishing (PIPS) to perforation using a Gatan 691 instrument. For planar samples the (See Figure~\ref{fig-foil-geometry}) the electron beam oriented normal to the shear plane, whereas for cross-sectional samples the electron beam is oriented parallel to this plane. Note that in both cases, electron beam is normal to the shear direction and elongation of the components will be visible.  PIPS made use of a liquid nitrogen-cooled sample stage to minimise specimen heating during foil preparation. The foils examined were extracted from the outer section of the HPT disc, and the hole position corresponds to a radial distance of approximately 3\,mm. TEM and scanning TEM (STEM) observations were made using  CM12 and JEOL Cs-corrected 2100F  microscopes operating at 120\,kV and 200\,kV respectively. High angle annular dark field STEM (HAADF-STEM) images were obtained on the JEOL microscope with inner and outer collection angles of 65.51 and 174.9\,mrad, respectively. Lamellar spacings were determining from the STEM images using the same procedure as for the SEM micrographs. It was not possible to obtain reliable data from heavily deformed Cu50Mo50 composites due to the decomposition of the lamellar structure, as is described in the following section.

\section{Results}

\subsection{Hardness testing}

\begin{figure}[hbtp]
	\begin{center}
	\hfill
	\subfigure[Step 1 HPT\label{fig-vhn-strain-stage1}]{\includegraphics[width=8cm]{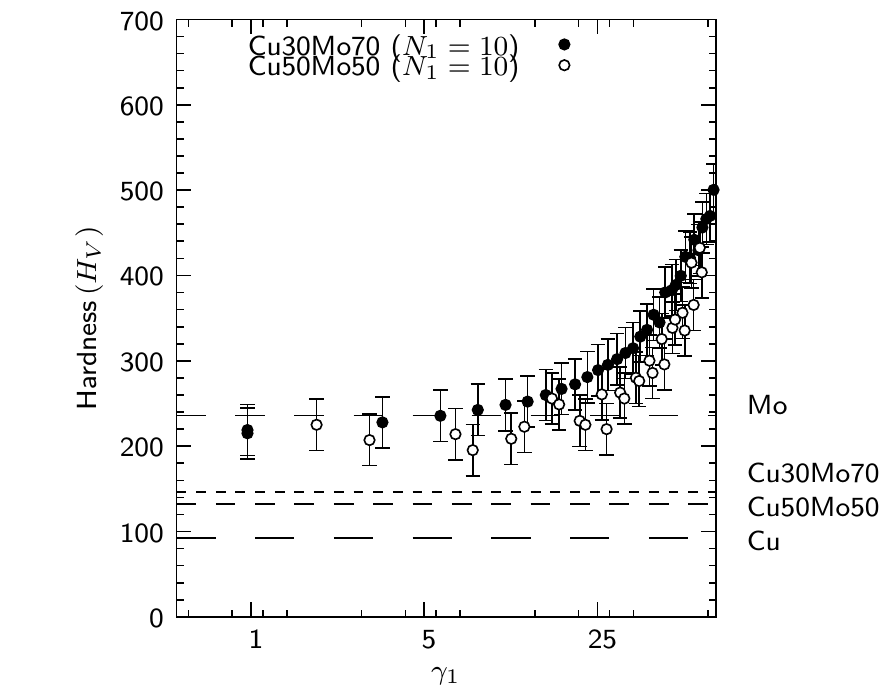}} 
	\hfill
	\subfigure[Step 2 HPT\label{fig-vhn-radius-stage2}]{\includegraphics[width=8cm]{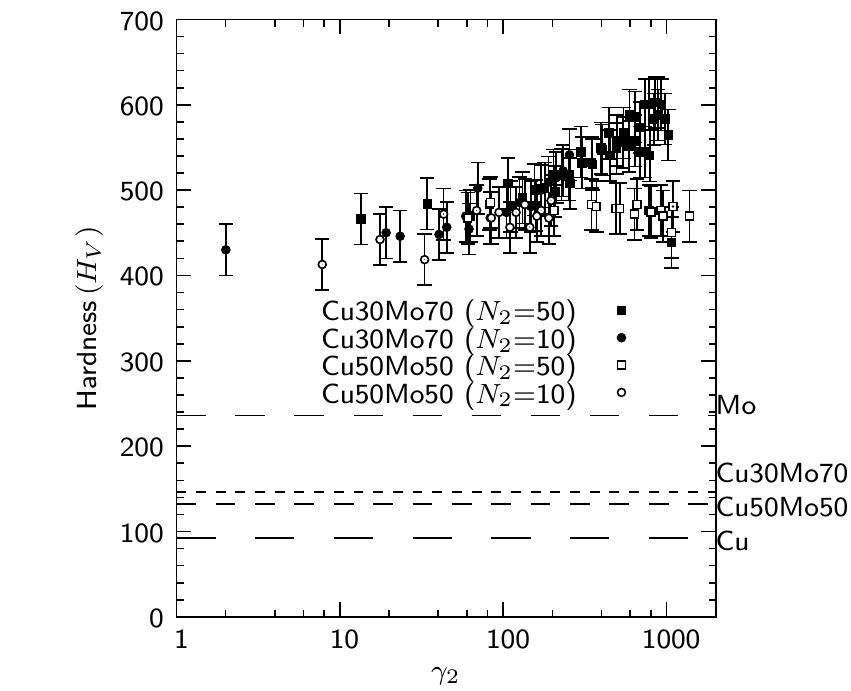}}
	\hfill\ 
	\caption{Vicker's indentation hardness of HPT Cu50Mo50 and Cu30Mo70 alloys as a function of shear strain after (a) step 1 and (b) step 2 HPT deformation. Strain values are indicated by $\gamma_1$ and $\gamma_2$, respectively. Hardness values for the as-received Cu30Mo70 and Cu50Mo50 alloys are shown, along with values measured for pure Cu and Mo.\label{fig-vhn-strain}}
	\end{center}
\end{figure}

The Vickers microindentation hardness after step one and step two HPT deformation is shown in Figures~\ref{fig-vhn-strain}(a) and (b), respectively. Values for the as-received Cu30Mo70 and Cu50Mo50 alloys are shown on the plot, along with values measured for pure Cu and Mo. The strain values are given as the von Mises strain (Eqn.~\ref{eqn-von-mises}) for each stage of deformation (i.e. $\gamma_1$ and $\gamma_2$, respectively). Assuming that slippage is negligible, the equivalent strain ($\epsilon_{eq}$) after $N$ rotations at a radial distance $r$ is given by, 
\begin{equation}
	\gamma_{eq}  \cong   \frac{ 2 \pi r N }{ \sqrt{3} t} 
\label{eqn-von-mises}
\end{equation}
where the disc thickness is $t$ \cite{CuberoSesin2012}.
(In the ideal case where co-deformation of both phases occurs, and slippage is absent, the strains in each step are multiplicative \cite{BachmaierKerber2012} giving a total equivalent strain of $\sim$50,000 for the maximum deformation used  in this work ($N_1$=10, $N_2$=50).

The hardness of both alloys increased rapidly during the first step of HPT deformation, to reach 400 $H_V$ in the case of Cu50Mo50 and 420 for Cu30Mo70 (Fig.~\ref{fig-vhn-strain-stage1}). The curves for both alloys are similar, and show the typical increase in hardness with applied strain (i.e. at greater distances from the centre of rotation), but with the Mo-rich composition showing a slightly higher hardness at equivalent levels of strain. 

Differences between the two compositions become more evident during step two HPT deformation. For Cu50Mo50 the least strained region (close to the neutral strain axis) showed no increase over the maximum value during step one deformation.
In the remainder of the disc the hardness increased to a plateau at 475\,$H_V$ after $N_2=10$ rotations, with no further increase when 50 rotations were applied. 

In contrast the Cu30Mo70 had a considerable increase in hardness during step 2 deformation with a substantial difference in hardness remaining between the highly strained edges and the lesser strained central region. The overall hardness  increased further for 50 rotations as compared to 10 rotations. After step two HPT (50 rotations) the hardness of  Cu30Mo70 ranged from  480\,$H_V$ to 600\,$H_V$. This was a substantial increase over the as-received condition, in which the hardness was measured as 146$\pm$4\,$H_V$ and pure Mo (236$\pm4H_V$). The wide range of hardness values and lack of a plateau region indicated that this material had not yet reached a strain-saturated condition.

\subsection{Scanning electron microscopy}

\begin{figure*}[hbtp]
\begin{center}

	\hfill
	Cu50Mo50
	\hfill
	Cu30Mo70
	\hfill\ 

	\hfill
	\subfigure[\label{fig-sem-hpt-inital1-Cu50Mo50_1}]{\includegraphics[width=6cm]{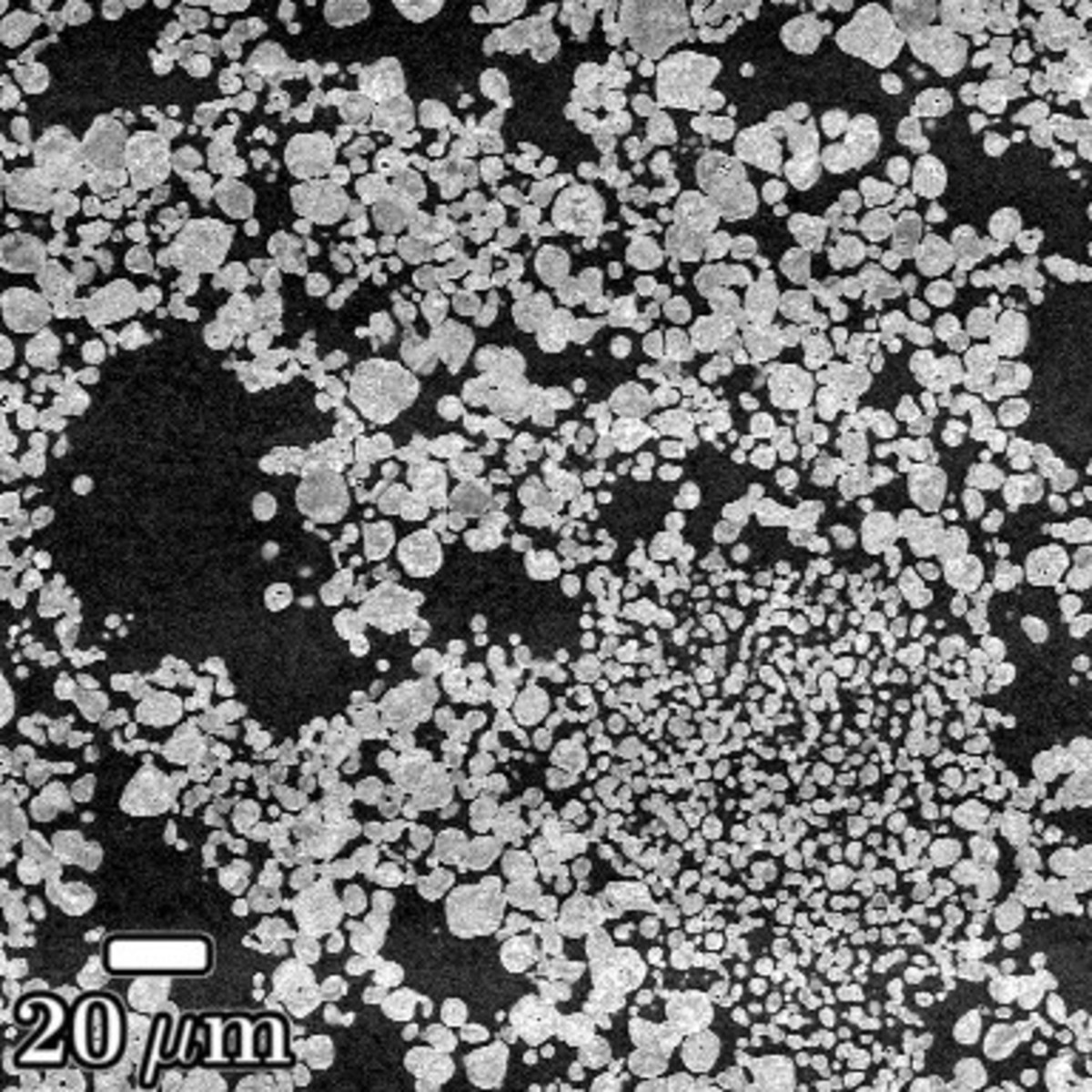}} 
	\hfill
	\subfigure[\label{fig-sem-hpt-inital1-Cu30Mo70_1}]{\includegraphics[width=6cm]{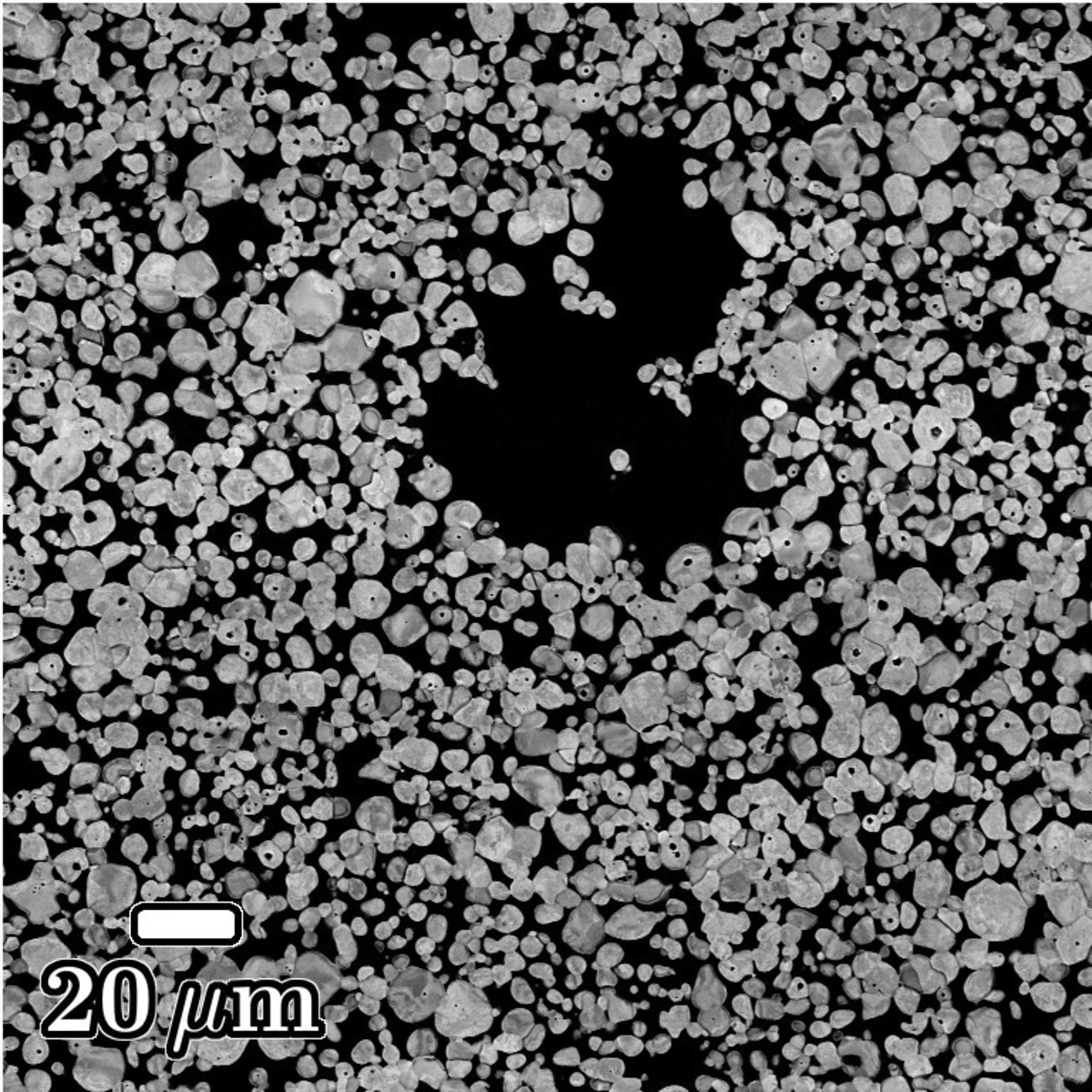}} 
	\hfill\ 

	\hfill
	\subfigure[\label{fig-sem-hpt-inital1-Cu50Mo50_2}]{\includegraphics[width=6cm]{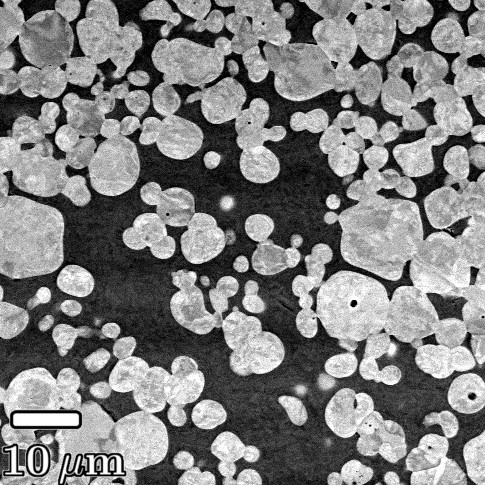}} 
	\hfill
	\subfigure[\label{fig-sem-hpt-inital1-Cu30Mo70_2}]{\includegraphics[width=6cm]{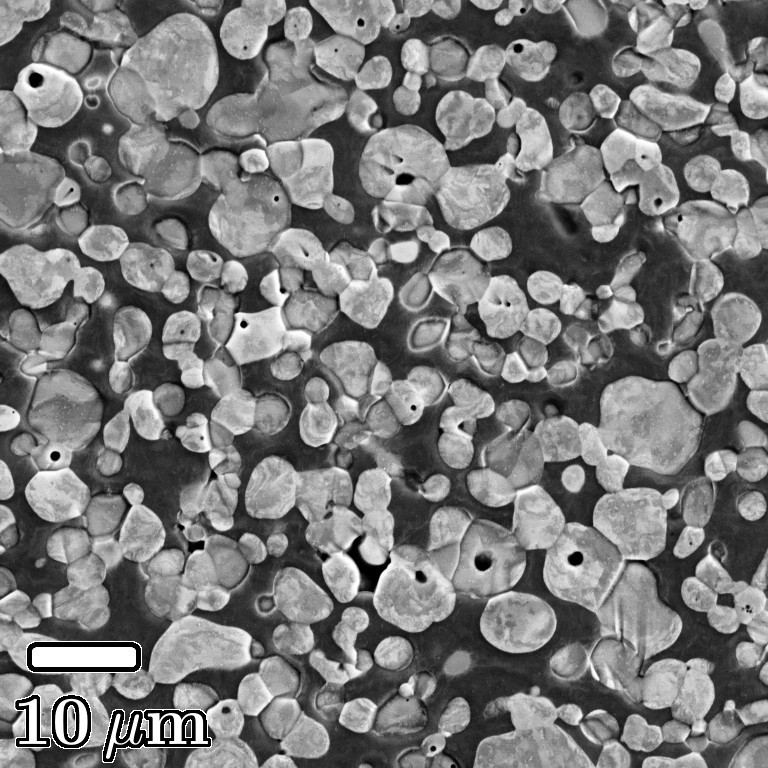}} 
	\hfill\ 

	\caption{Backscattered electron micrographs of as-received  (a,c) Cu50Mo50 and (b,d) Cu30Mo70 showing the interconnected network of Mo particles infiltrated by copper.  \label{fig-sem-hpt-inital1}}
\end{center}
\end{figure*}

\begin{figure}
	\begin{center}
	\includegraphics{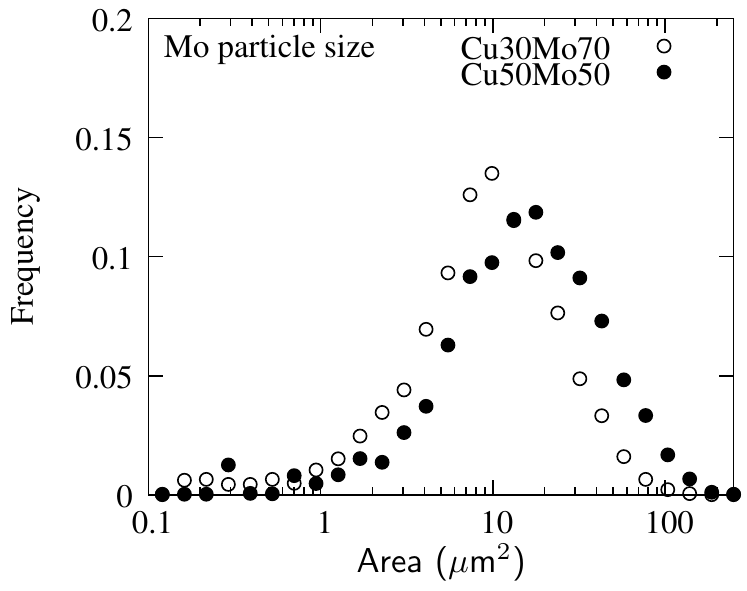}
	\caption{Stereological meaurements of the projected area of the Mo particles in the as-received composites.
\label{fig-stereol-initial}}
	\end{center}
\end{figure}

Scanning electron micrographs of the as-received composites showed a network of connected, spheroidal Mo particles. The micrographs shown were obtained using backscattered electron mode in which Cu (atomic no., Z=29) appears dark and Mo (Z=42) appears bright. 
In each of the micrographs the shear plane is horizontal and lies normal to the plane of the page. The main distinction between the two compositions was a higher number of  Cu-filled pores with diameters of 10-50\,$\mu$m diameter in the Cu50Mo50 alloy (Fig.~\ref{fig-sem-hpt-inital1-Cu50Mo50_1},\ref{fig-sem-hpt-inital1-Cu50Mo50_2}).  Voids of approximately 1\,$\mu$m in diameter are apparent in several of the Mo particles in the higher magnification images shown in Figures \ref{fig-sem-hpt-inital1-Cu50Mo50_2} and \ref{fig-sem-hpt-inital1-Cu30Mo70_2}. 

Stereological measurements were made to determine the distribution of  projected areas of the Mo particles, as shown in Figure~\ref{fig-stereol-initial}. Assuming a spherical particle geometry, the data yielded average diameters of 3.8\,$\mu$m for Cu30Mo70 and 5.0\,$\mu$m for Cu50Mo50. The particles were relatively monodisperse; in the case of Cu30Mo70,  50\% of Mo particles had diameters between 2.3 and 4.3\,$\mu$m. For the Cu50Mo50 composite the equivalent values were 2.8 and 5.7\,$\mu$m. 

\begin{figure*}
	\begin{center}
	\hfill
	Cu50Mo50
	\hfill
	Cu30Mo70
	\hfill\

	\hfill
	\rotatebox{90}{\parbox{6cm}{\centering Rim}}
	\subfigure[]{\includegraphics[width=6cm]{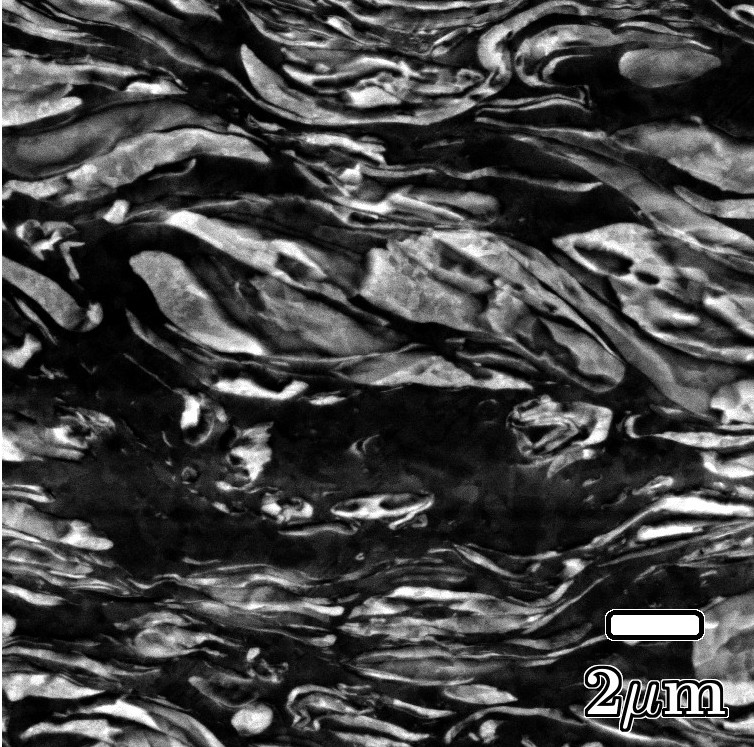}} 
	\hfill
	\subfigure[]{\includegraphics[width=6cm]{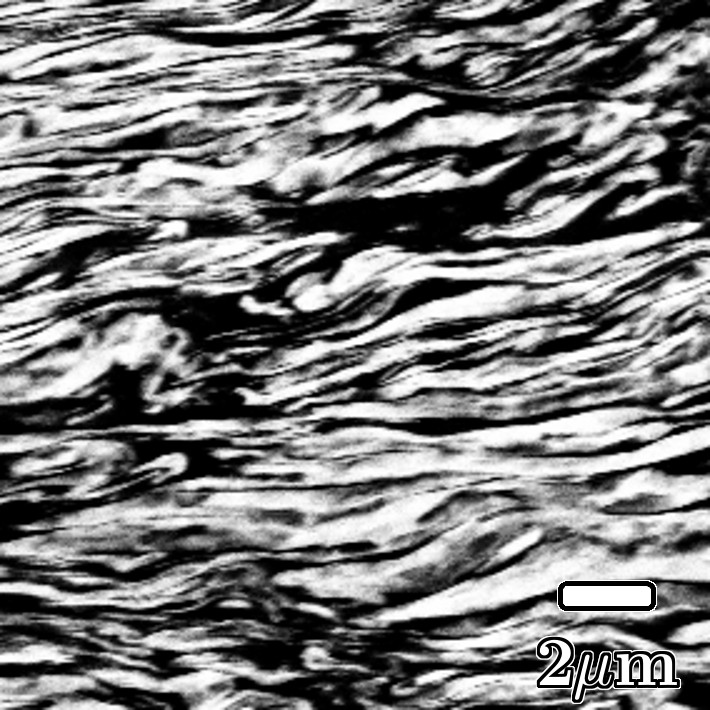}} 
	\hfill\

	\hfill
	\rotatebox{90}{\parbox{6cm}{\centering Near-centre}}
	\subfigure[]{\includegraphics[width=6cm]{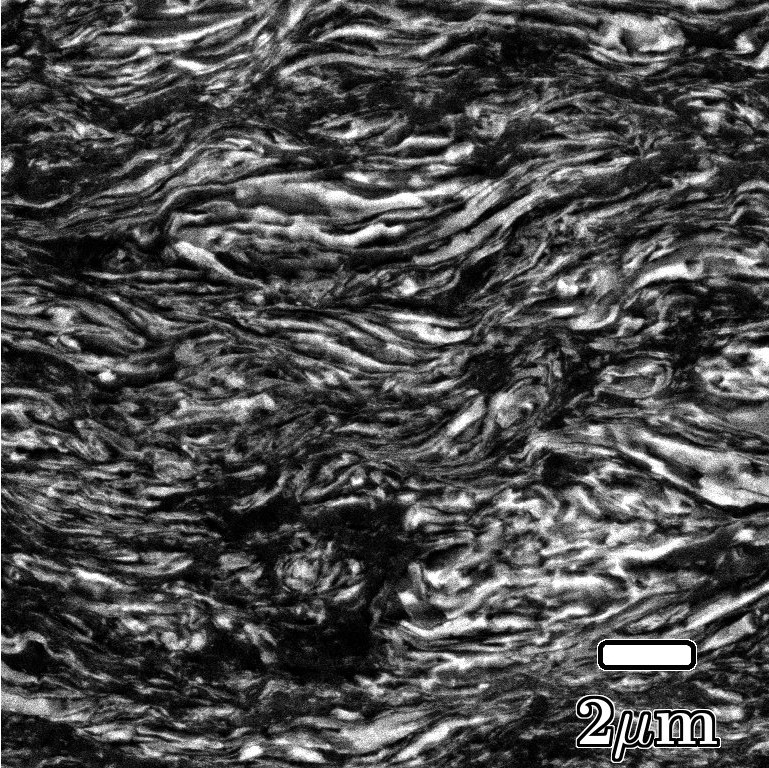}} 
	\hfill
	\subfigure[]{\includegraphics[width=6cm]{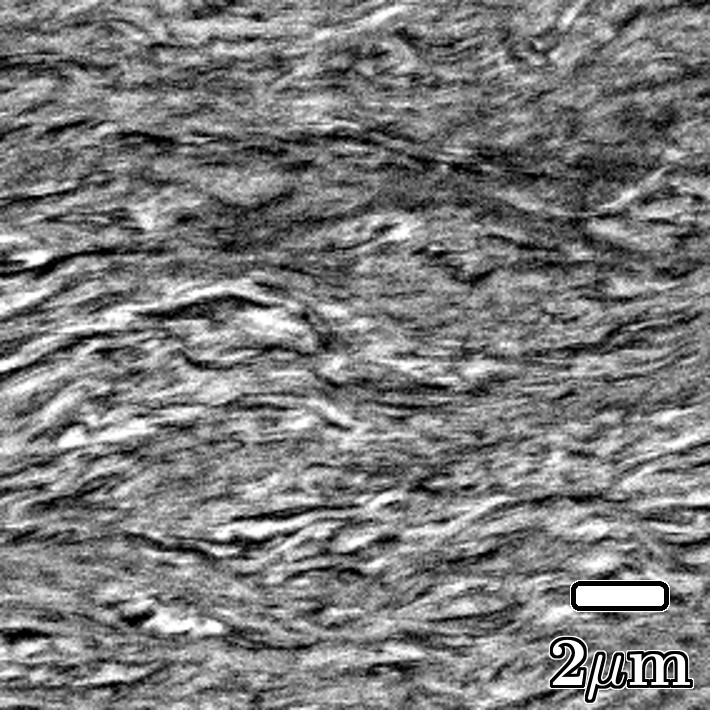}} 
	\hfill\

	\caption{Backscattered electron micrographs of (a,c) Cu50Mo50 and (b,d) Cu30Mo70 showing the lamellar structure after step one HPT deformation (10 rotations, $\gamma_1\approx70$). Mo (atomic number, Z, =42) appears bright, with Cu-containing (Z=29) regions appearing dark. \label{fig-sem-hpt-stage1}}

\end{center}

\end{figure*}

A lamellar microstructure began to develop during step one HPT deformation.
Figure~\ref{fig-sem-hpt-stage1} presents typical micrographs obtained at this step of deformation, with the shear direction in the horizontal plane of the page. The development of a lamellar microstructure can be seen by comparing images recorded at the centre of the samples (a,c) where the local strain is minimal (and theoretically zero at the exact rotation axis) and those recorded at the outer rim of the sample (b,d) where the maximum strain was imposed. The lamellae were finer in the Mo-rich composition and could  be identified at lower strain. For example, a lamellar structure is already visible at the centre of the Cu30Mo70 composite (c) but not in Cu50Mo50 (a), although some co-deformation is evident. 

Backscattered electron micrographs of samples given step 2 deformation showed uniform brightness with no lamellae or individal grains being identified. The absence of well-defined changes in atomic contrast was taken as evidence that either a) a single-phase microstucture had developed of b) that microstructure was of a finer scale than the interaction volume of the SEM could resolve. This required recourse to transmission electron microscopy to determine and is discussed in the following section.

\subsection{Transmission electron microscopy}

\begin{figure*}[hbtp]
	\hfill
	\subfigure[\label{fig-tem-overview}]{\includegraphics[width=6cm]{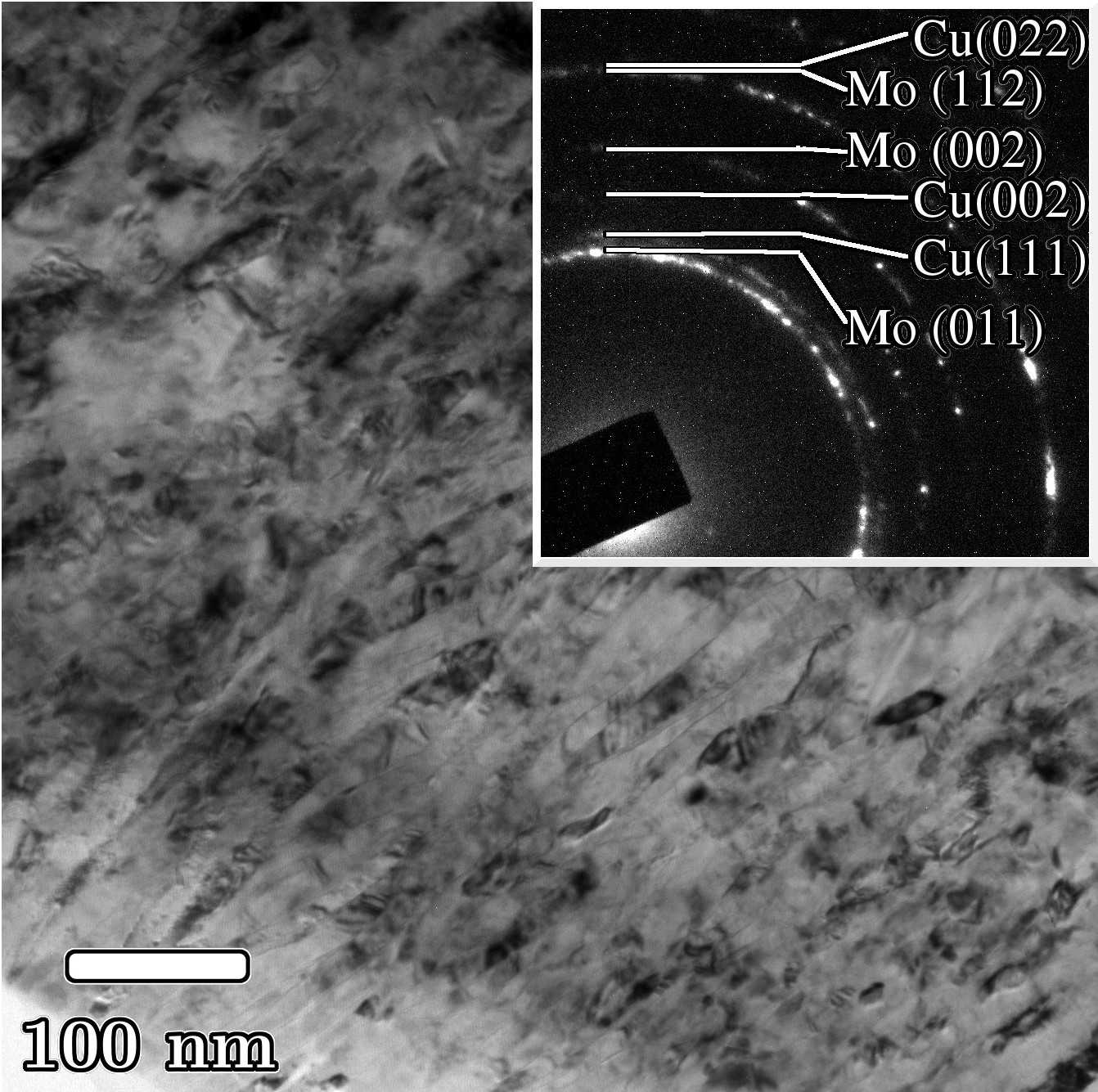}}
	\hfill\

	\hfill
	\subfigure[\label{fig-hrtem-cu30-hpt1050-1}]{\includegraphics[width=6cm]{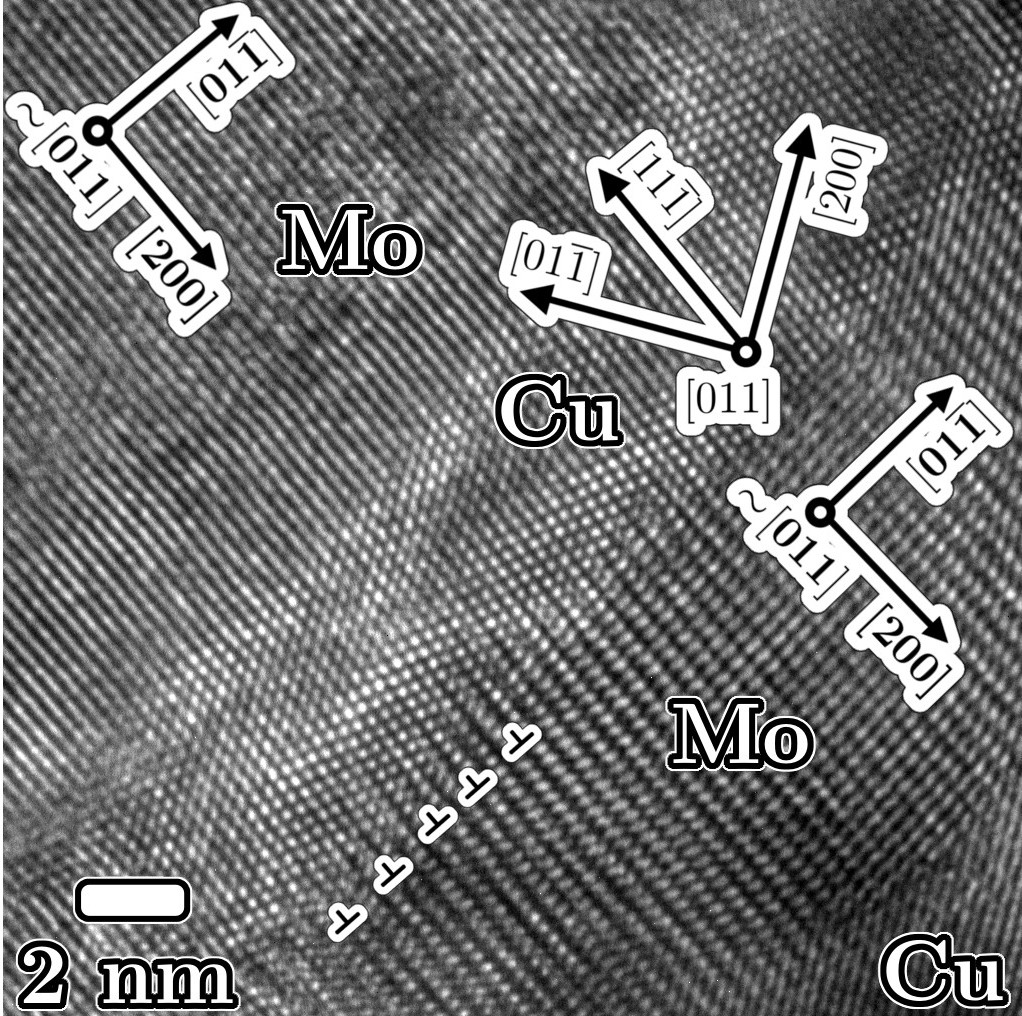}}
	\hfill\

	\caption{TEM images of Cu30Mo70 after two step HPT deformation ($\gamma_2\approx750$). (a) shows a bright field, diffraction contrast image giving an overview of the microstructure. (b) shows a high-resolution image showing alternating copper and molybdenum lamellae.}
\end{figure*}

TEM micrographs of Cu30Mo70 subjected to two-step deformation show a microstructure comprised of lamellae with a thickness of  5--20\,nm.   Figure~\ref{fig-tem-overview} shows a typical region of a foil of Cu30Mo70 after two-step HPT deformation ($N_1$=10, $N_2=50$) corresponding to equivalent strains $\gamma_1\approx50$, $\gamma_2\approx750$.  
The shear plane in both micrographs is angled from top right to bottom left and lies normal to the plane of the page. 
The selected area diffraction pattern (inset) show reflections attributable to both phases. 
Figure~\ref{fig-hrtem-cu30-hpt1050-1} shows a high-resolution TEM image showing a series of alternating Cu and Mo lamellae. The central Cu region is viewed along the [011] zone axis and  both the adjacent Mo lamellae are close to this orientation. 

\begin{figure*}[hbtp]
	\begin{center}
	\hfill
	BF
	\hfill
	HAADF
	\hfill\ 

	\hfill
	\subfigure[\label{fig-stem-cu30-hpt1050-1}]{\includegraphics[width=6cm]{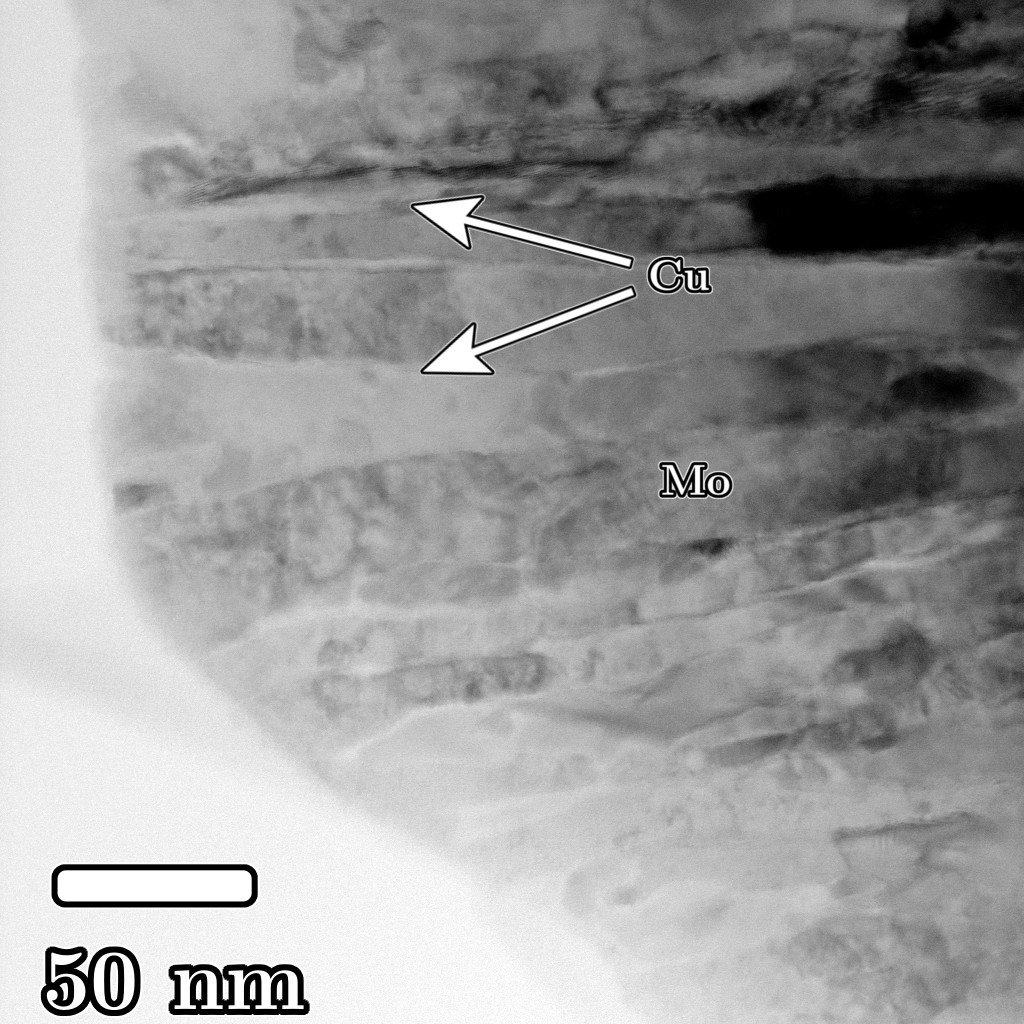}}
	\hfill
	\subfigure[\label{fig-stem-cu30-hpt1050-2}]{\includegraphics[width=6cm]{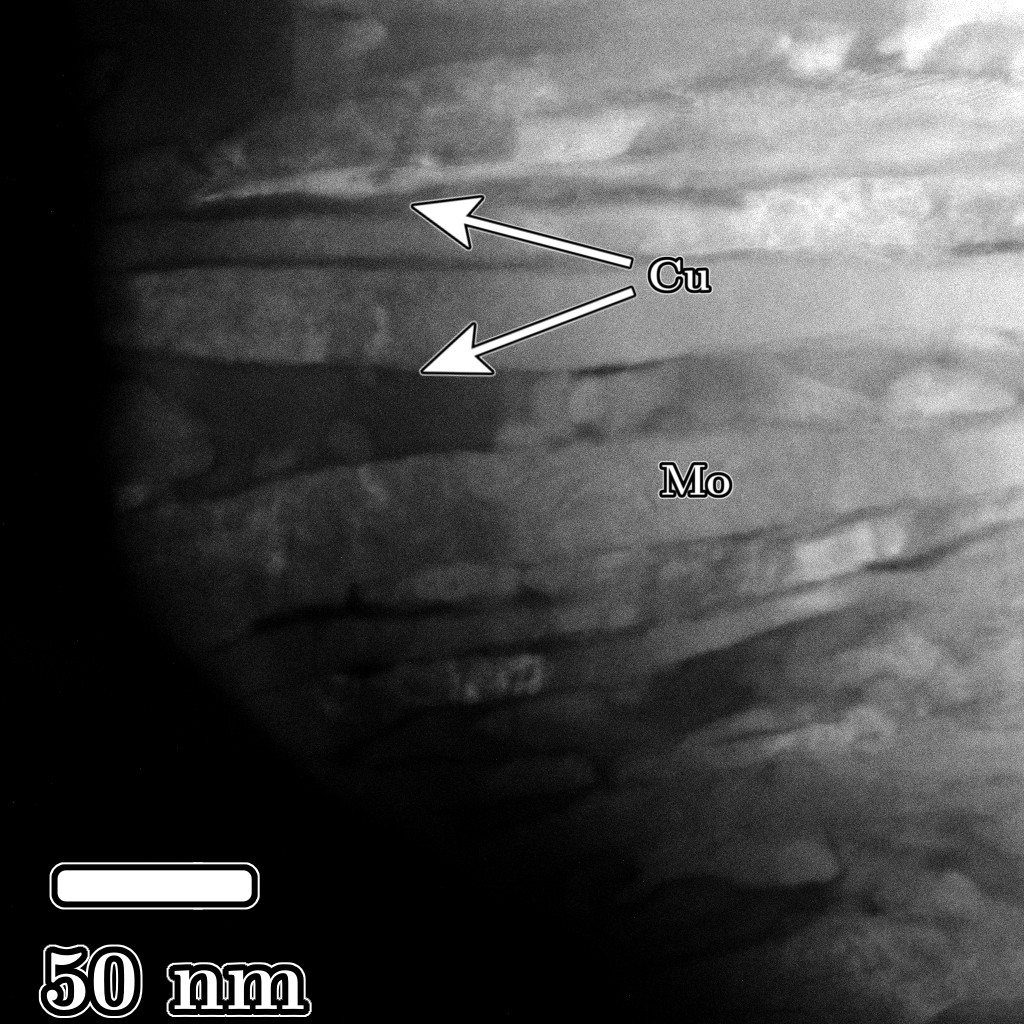}}
	\hfill\

	\hfill
	\subfigure[\label{fig-stem-cu30-hpt1050-3}\label{fig-grain-size-bf}]{\includegraphics[width=6cm]{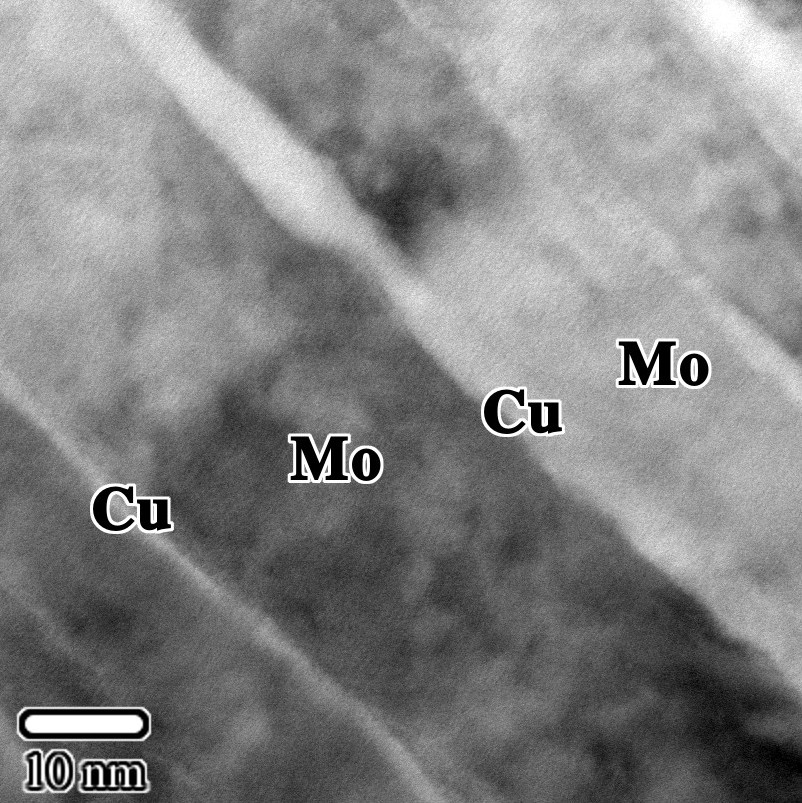}}
	\hfill
	\subfigure[\label{fig-stem-cu30-hpt1050-4}\label{fig-grain-size-cdf}]{\includegraphics[width=6cm]{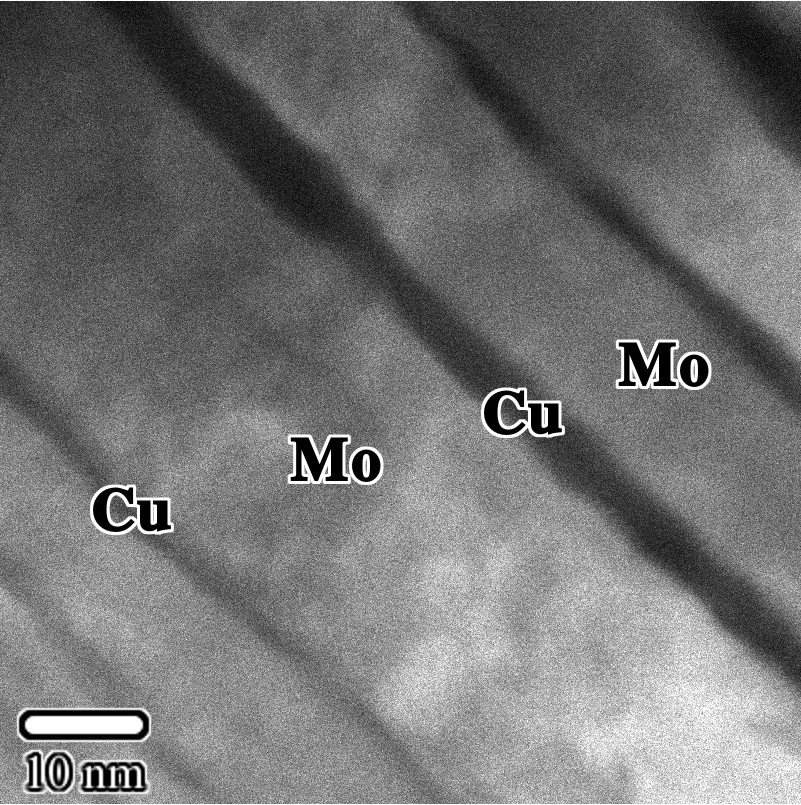}}
	\hfill\
	\caption{(a,c) Bright-field and (b,d) HAADF STEM images of Cu30Mo70 after two-step HPT deformation ($\gamma_2\approx750$). Mo-rich regions appear bright in (b,d) \label{fig-stem-cu30-hpt1050}}
	\end{center}
\end{figure*}

Mo- and Cu-rich regions are readily distinguished in High Angle Annular Dark Field (HAADF) Scanning Transmission Electron Microscope (STEM) images.  Mo lamellae (bright regions in the the micrograph) were 10--20\,nm in thickness, whereas the (darker) Cu lamellae were generally   $\sim$5\,nm in thickness. Figure~\ref{fig-stem-cu30-hpt1050} presents a series of typical bright field and HAADF-STEM images of Cu30Mo70 after step two HPT deformation. Both Cu and Mo lamellae are highly elongated, with lengths of 200\,nm or longer and are aligned closely with the shear plane which is normal to the plane of the page and is oriented horizontally in (a,b) and from top left to bottom right in (c,d)). Stereological measurements of the lamellar spacing using circular grids gave a lamellar spacing of 9\,nm.

\begin{figure*}[hbtp]
	\begin{center}
	\hfill
	BF
	\hfill
	HAADF
	\hfill\ 

	\hfill
	\subfigure[\label{fig-stem-cu50-hpt1050-1}]{\includegraphics[width=6cm]{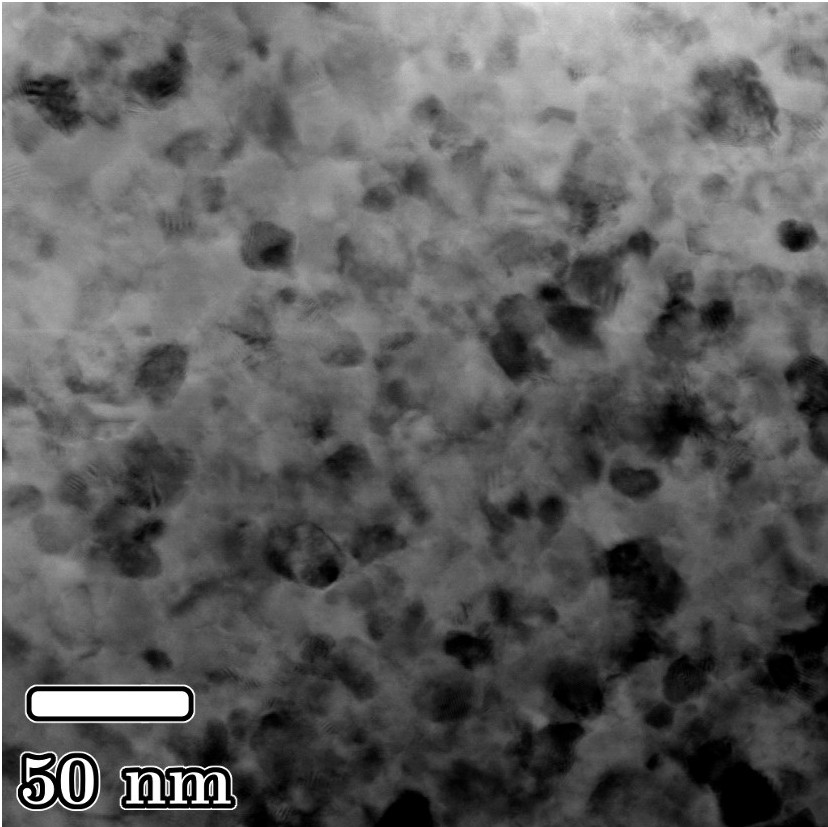}}
	\hfill
	\subfigure[\label{fig-stem-cu50-hpt1050-2}]{\includegraphics[width=6cm]{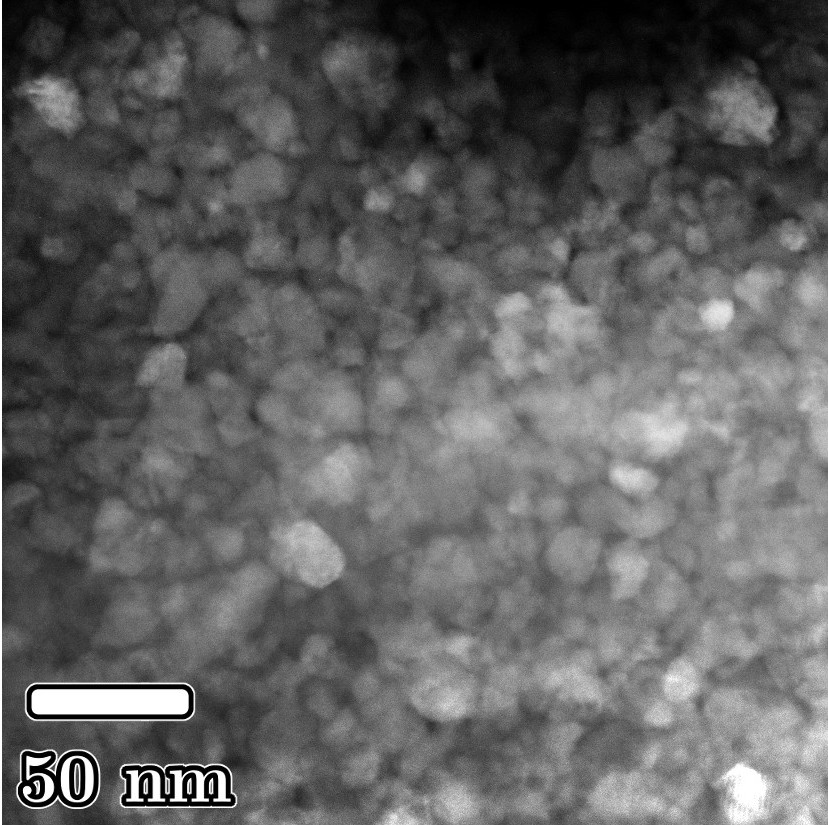}}
	\hfill\ 

	\hfill
	\subfigure[\label{fig-stem-cu50-hpt1050-3}\label{fig-grain-size-bf}]{\includegraphics[width=6cm]{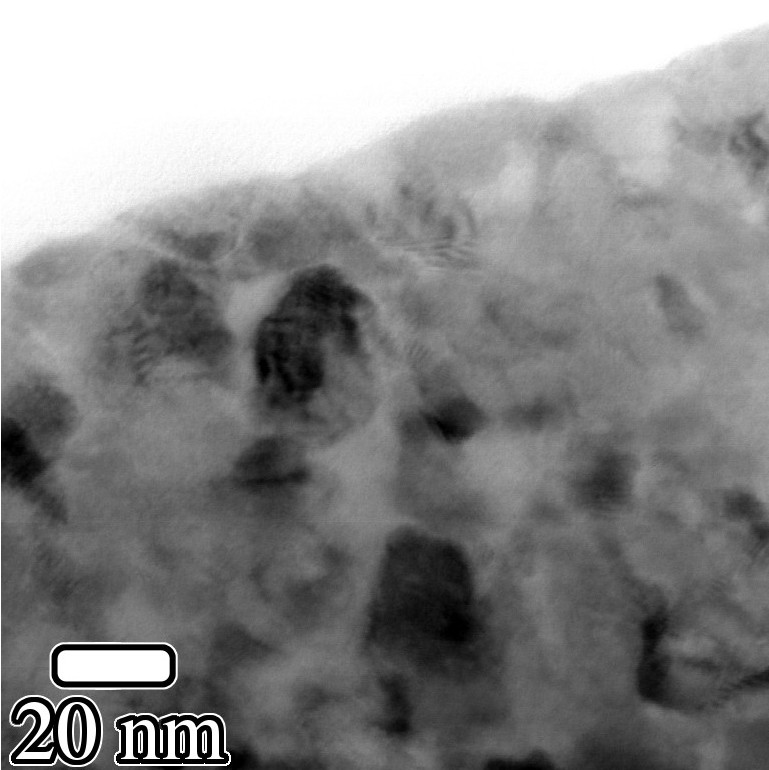}}
	\hfill
	\subfigure[\label{fig-stem-cu50-hpt1050-4}\label{fig-grain-size-cdf}]{\includegraphics[width=6cm]{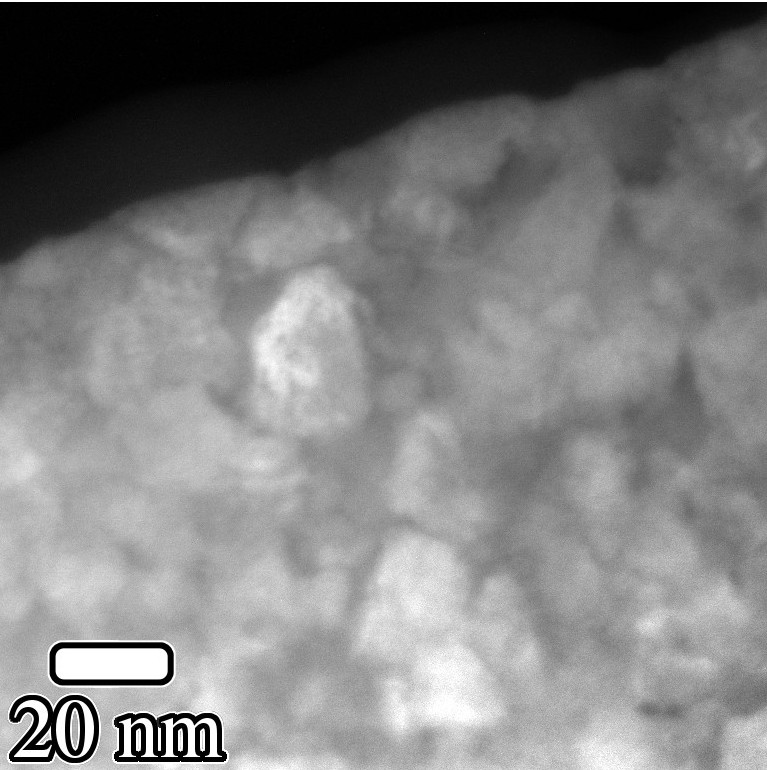}}
	\hfill\ 

	\caption{(a,c) Bright-field and (b,d) HAADF STEM images of Cu50Mo50 after two-step HPT deformation ($\gamma_2$=750). Mo-rich regions appear bright in (b,d) \label{fig-stem-cu50-hpt1050}}
	\end{center}
\end{figure*}

The microstructure of the Cu50Mo50 alloy did not remain lamellar after the second step of HPT deformation. HAADF-STEM micrographs of Cu50Mo50 show  equiaxed Mo grains with diameters of approximately 10--15\,nm (Fig~\ref{fig-stem-cu50-hpt1050}).  The Mo particles (appearing as darker regions in (a,c) and bright regions in the HAADF-STEM images (b,d)) appear to be embedded within the Cu component, with a separation of $<$5\,nm between adjacent Mo particles. There appeared to be no relationship between the particle morphology and the shear plane which lies in the horizontal plane.

\section{Discussion}

Liquid-metal infiltrated Mo-Cu proved to be amenable to room-temperature deformation via HPT, making it possible to produce   nanometer-scale composites.  No radial cracking was observed after step one deformation, where the maximum equivalent strain was $\sim72$. Although radial cracks were sometimes evident after step two deformation, it was generally possible  to deform to 50 revolutions. Deformation-induced heating during HPT is only  expected to raise the temperature to around 373K during step one deformation and 300K during step two deformation. This is much lower than the temperatures of 573-673\,K\cite{ShieldsLipetzky2001} usually required to ensure sufficient ductility in Mo,  however, the high imposed pressure and quasi-hydrostatic conditions allowed for deformation of both Cu and Mo components. 

Due to the fine scale of the HPT material and the absence of intergranular porosity, the hardness values in the strain-saturated state compare very favourably with densified ball-milled materials. The hardness values reported in Mo-Cu with Mo $<20$ wt.\% alloys were in the range of 200--259 $H_V$ \cite{KumarJayasankar2015,SabooniMousavi2012,AguilarCastro2012} and ball-milled and annealed pure Mo was reported to have a hardness of 250 $H_V$. These value are substantially below the values of 475\,$H_V$ and 600\,$H_V$ measured herein for HPT-deformed Cu50Mo50 and  Cu30Mo70, respectively.

Both of the  Cu-Mo composites developed similar lamellar microstructures during the first step of deformation (Fig.~\ref{fig-sem-hpt-stage1}), with the microstructure being more refined at higher strains (c,d) and in the Cu30Mo70 composite for equal strain (b,d). However, in the Cu50Mo50 composite this microstructure decomposes during  step two deformation in  Cu50Mo50 (Fig~\ref{fig-stem-cu50-hpt1050}) unlike Cu30Mo70(Fig~\ref{fig-stem-cu30-hpt1050}). 
 Previous investigations on binary systems containing phases with substantial differences in hardness have reported the lamellar structure formed during HPT breaks up at higher strains and is replaced by a microstructure consisting of equiaxed grains \cite{BachmaierRathmayr2014}. It has been argued that once such microstructures develop, additional deformation is localised in the softer phase and does not further refine the microstructure \cite{BachmaierRathmayr2014}.  The microstructure then enters a steady state, with the mechanical properties, including hardness, remaining unchanged even if further strain is applied. Strain-localisation in the soft phase occurs more readily for small volume fractions of the hard phase, where there is limited contact between particles of the hard phase during deformation. Therefore the higher volume fraction of Mo in Cu30Mo70 (67\% compared to 47\% in Cu50Mo50) is thought to be responsible for the retention of a lamellar structure even after extensive two-step deformation. 

\begin{figure}
	\begin{center}
	\includegraphics[width=7cm]{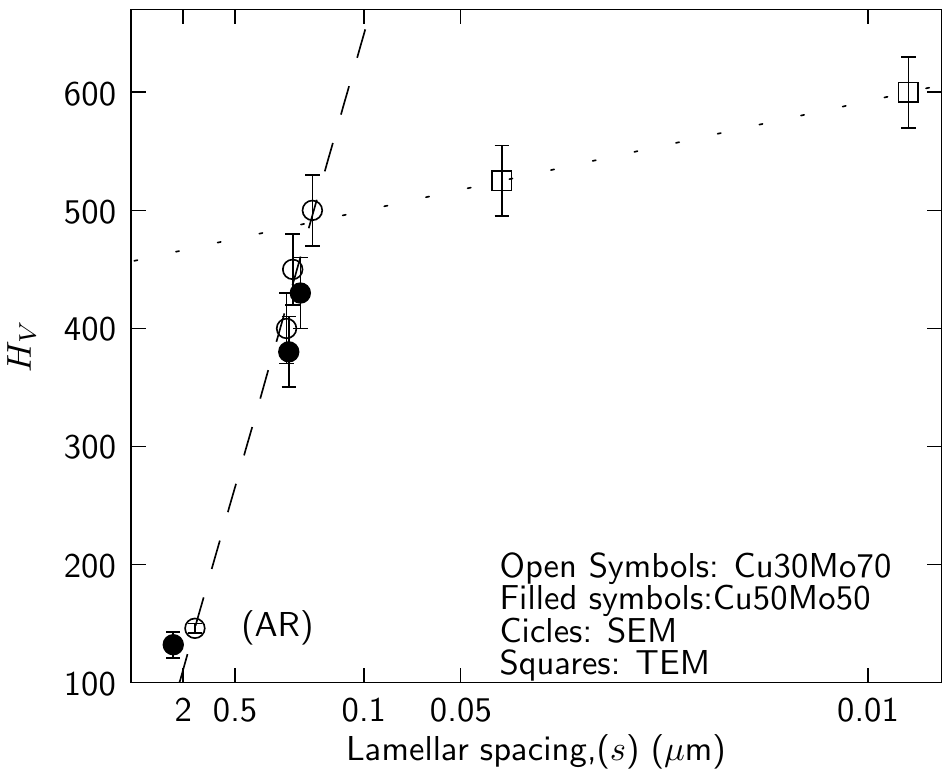}
	\caption{Hardness of the Cu-Mo composites plotted against the lamellar spacing,$s$.\label{fig-hall-petch}
A Hall-Petch co-efficient of $K=217\pm7$\,MPa/$\mu$m$^{1/2}$ was determined from the gradient for $s>160$\,nm. 
The line for $s<160$\,nm is provided as a guide for the eye, only. }
	\end{center}
\end{figure}

The relationship between the hardness and lamellar spacing, ($s$), of the composites is presented in Figure~\ref{fig-hall-petch}, which shows a Hall-Petch dependence on the lamellar spacing for spacings as low as 160\,nm.  The gradient of the hardness values plotted against $1/\sqrt{s}$ for lamellar spacing, $s$, gave a Hall-Petch coefficient, $K$, of $217\pm7$\,MPa/$\mu$m$^{1/2}$. Similar relationships between lamellar spacing and hardness were demonstrated in eutectoid pearlitic steels \cite{Ray1991} and have also been established for the widely-studied Cu-Nb composites\cite{TrybusSpitzig1989,ThillyLecouturier2002,Spitzig1991,CarpenterZheng2013}.  This behaviour indicates that plastic deformation is interface controlled, due to the difficulty of transfering dislocations across the interface. However, further refinement of the microstructure had little effect, with a factor of 20 reduction in the spacing leading to a hardness increase of approxmately 20\%. 
This deviation from the Hall-Petch relationship for extremely fine structures has been reported in nanoscale Cu-Nb alloys \cite{MitchellLu1997}. This is thought to indicates a change in the deformation mechanism, with studies suggesting that interface sliding occurs\cite{WangHoagland2011}.

The microstructural development of the Cu-Mo composites is initially similar to that reported for HPT-deformation of other Group VI refractory metals. HPT deformation of a W-25\,wt.\%Cu composite resulted in a lamellar microstructure  for equivalent strains of at least 64 \cite{SabirovPippan2005}. Another study using a two-step HPT procedure subjected W-25wt.\%Cu to equivalent strains of 1000 without reaching a hardness plateau \cite{KraemerWurster2014}. This would suggest that the material retains a lamellar microstructure under these conditions. HPT Cu-Cr composites have also been examined at  high strains \cite{BachmaierRathmayr2014,GuoRosalie2017} and it has been shown that the lamellar structure decomposed to form equiaxed, roughly spheroidal particles after strains of $\sim$250\cite{GuoRosalie2017}. These reached a grain size of $\sim15$\,nm after strain of $\approx1400$ but showed no further reduction in grain size at a strain of $\approx4000$\cite{GuoRosalie2017a}. In comparison, an average equivalent shear strain of $\sim$50 was imposed on the Cu-Mo composites duing the first HPT deformation step, and a further (multiplicative) strain of $\sim$1000 during the second deformation step, indicating the the Cu30Mo70 layered microstructure is stable to extremely high strains.

The deformability of the refractory phase at a given temperature and hence the ductile to brittle transformation temperature are also critical to the retention of the lamellar microstructure. The DBTT of tungsten (370-470\,K\cite{GumbschRiedle1998}) is 
only slightly higher than the value of $\sim$373\,K \cite{ShieldsLipetzky2001} for Mo. However, chromium has a DBTT variously reported between 423--773K\cite{WadsackPippan2002,HolzwarthStamm2002,LuconVanWalle2001} and in Cu-Cr composites the lamellar structure decomposes at much lower strains than either Cu-W or Cu-Mo. 

Composites with similar layer thicknesses of $\sim$10--20\,nm layers have also produced via accumulative roll-bonding (ARB) of pure Cu and Nb \cite{Beyerlein2013,EkizLach2014}. These layered structures were retained during subsequent HPT deformation of the ARB composite\cite{EkizLach2014}. It is important to note, however, that molybdenum is significantly less ductile than niobium and that the quasi-hydrostatic conditions during HPT play a role in allowing for Mo/Cu co-deformation at room temperature. HPT therefore offers a means of generating ultrafine layered structures containing less ductile refractory elements, opening up the possibility of more detailed studies of their microstructure and properties. One area of interest concerns the texture of the composites, since the ARB Cu-Nb composites evolved atypical textures as the layer thickness approached the grain size due to the increasing influence of interfaces and the constraint imposed by neighbouring grains \cite{Beyerlein2013}. Further work is underway to examine the texture of Cu-Mo HPT composites.

\section{Conclusions}
	Liquid-metal infiltrated Mo-Cu composites were found to be amenable to high-pressure torsion at room temperature. Both copper and molybdenum phases underwent deformation giving a lamellar microstructure after the first step of HPT deformation, corresponding to an average equivalent shear strain of 50. In Cu50Mo50 the lamellar structure breaks up during the second step of HPT deformation and the microstructure consisted of equiaxed Mo nanoscale grains surrounded by Cu. This behaviour is similar to that of Cu-Cr composites and liquid metal infiltrated Cu-W composites, and is associated with a strain-saturated state in which the hardness reaches a plateau. However, in the Cu30Mo70 material the lamellar structure was retained and after stage-two deformation the resulting composites had a fine lamellar structure with Cu and Mo forming 5\,nm and 10-20\,nm lamellae respectively, with an overall lamellar spacing of 9\,nm. Notwidthstanding the poor room-temperature ductility of Mo, this  microstructure was stable to equivalent strains of $\sim$50,000 and the layer thicknesses became equivalent to those of accumulative roll-bonded Cu-Nb. Since no sintering was used, post-deformation grain growth was avoided. The fine microstructure and absence of porosity contributed to Vickers hardness values of 475 for Cu50Mo50 and 600 for Cu30Mo70. The combination of high strength, and a fine, oriented microstructure would be well-suited to thermoelectric materials and show that  HPT-deformed Cu-Mo composites warrant further investigation for such applications.

\section*{Acknowledgements}
This work was conducted under FWF project 27034-N20 ``Atomic resolution study of deformation-induced phenomena in nanocrystalline materials''.  The starting materials were provided by Plansee, Austria. SEM sample preparation was performed by S. Modritsch and some of the TEM sample preparation by G. Felber. Some of the hardness testing was carried out by I. Lukanovic. The authors are grateful for valuable discussions with P. Ghosh and O. Renk.

\newcommand{\noopsort}[1]{} 
  \newcommand{\singleletter}[1]{#1} \newcommand{\switchargs}[2]{#2#1}
  \newcommand{\printfirst}[2]{#1}

\end{document}